\documentclass[journal]{new-aiaa}
\usepackage[utf8]{inputenc}

\usepackage{graphicx}
\usepackage{amsmath}
\usepackage[version=4]{mhchem}
\usepackage{siunitx}
\usepackage{longtable,tabularx}
\usepackage{booktabs}
\usepackage{pgfplots}

\usepackage{fancyhdr}
\title{Hybrid RANS-LES simulation of transverse fuel injection in a Mach-10 scramjet engine} 

\author{Nick Plewacki\footnote{CFD Engineer, Survice Engineering, APG, MD 21005, and AIAA Member}, Benjamin Kale\footnote{Graduate Researcher, Dept. of Mechanical Engineering, University of Maryland Baltimore County}, Manu Kamin\footnote{Research Associate, Dept. of Aerospace Engineering, University of Cincinnati}, Luis Bravo\footnote{Sr. Aerospace Engineer, DEVCOM Army Research Laboratory, 6340 Rodman Road, APG, MD 21005, and AIAA Associate Fellow. Corresponding author: \href{mailto:luis.g.bravorobles.civ@army.mil}{luis.g.bravorobles.civ@army.mil}}\\Survice Engineering, Aberdeen Proving Ground, MD 21005, \\Mechanical Engineering Department, University of Maryland Baltimore County, MD, 21250 \\Aerospace Engineering Department, University of Cincinnati, OH, 45220\\DEVCOM Army Research Laboratory, Army Research Directorate, Aberdeen Proving Ground, MD 21005}

\begin{document}

\maketitle

\begin{abstract}
Hypersonic flight poses unique propulsion challenges, requiring engines that maintain thrust, efficiency, and stability across a wide range of operating conditions. These engines must transition smoothly between flight regimes and altitudes. Scramjets (supersonic combustion ramjets) play a key role in addressing these challenges. Recent advancements in high-fidelity computational fluid dynamics (CFD) tools allow researchers to explore novel designs and improve the feasibility of hypersonic travel. In this work, we analyze a radical-farming type scramjet engine mounted at the University of Queensland's T4 Wind Tunnel at Mach 10. We use the Improved Delayed Detached Eddy Simulation (IDDES) model, which combines Reynolds-Averaged Navier–Stokes (RANS) and Large Eddy Simulation (LES) in different flow regions. A novel integrated modeling strategy is introduced, coupling the inlet, fuel injectors, combustor, and nozzle for full-scale engine analysis. Hydrogen combustion is modeled using a Finite Rate Chemistry (FRC) approach with a 12-species, 27-reaction mechanism to capture shock-induced chemical kinetics across equivalence ratios of $\phi = 0.5$ to $0.9$. The Takeno flame index analysis reveals multiple combustion regimes, with ignition occurring in the partially premixed regime. This is supported by Chemical Explosive Mode Analysis (CEMA), which identifies regions of high chemical sensitivity, correlating with observed hot pockets and providing insights into autoignition and flame stabilization mechanisms. The combination of IDDES and FRC improves the transport of hydrogen to hot pockets, producing combustion patterns that match experimental results. This work establishes a framework to address critical challenges in future air-breathing propulsion systems.
\end{abstract}


\section{Nomenclature}
{\renewcommand\arraystretch{1.0}
\noindent\begin{longtable*}{@{}l @{\quad=\quad} l@{}}
$M_{\infty}$  & Mach number \\
$Re_{\infty}$ & Reynolds number \\
$RANS$  & Reynolds Averaged Navier Stokes \\
$IDDES$ & Improved Detached Delayed Eddy Simulation \\
$C_{D}$ & Drag coefficient \\
$C_{L}$  & Lift coefficient \\
$C_{p}$  & Surface pressure coefficient \\
$Z_{FO}$   & Mixedness \\
$G_{FO}$   & Takeno Flame Index \\
$CEMA$   &  Chemical Explosive Mode Analysis\\

\newpage 

\end{longtable*}}

\section{Introduction}
\label{sec:intro}
\vspace{5mm}
Scramjet engines, while having been the subject of extensive research for decades, have gained renewed attention as a promising technology for long-range hypersonic flight and access to low Earth orbit. Despite notable advancements in the design and operation of scramjet engines, several fundamental challenges continue to hinder their practical implementation \cite{GONZALEZ_pecs_2017, urzay_arfm_2018, JOFRE_pecs_2021}. Among these challenges, the issue of inadequate fuel–air mixing is particularly critical and warrants thorough examination. Fuel-air mixing in a scramjet combustor occurs in a dynamic environment involving the complex interplay  between transverse fuel injection, shock wave interactions, and turbulent boundary layer dynamics. It is essential to obtain a homogeneous fuel–air mixture to achieve stable combustion reactions, reliable thrust and high propulsion efficiency over a wide range of flight Mach numbers $(M_\infty = 5 \text{--} 25) \text{ and altitudes (20--55 km)}$. However, under the conditions prevalent during supersonic flight, this task becomes increasingly complex due to the exceedingly short residence times of fuel and air (on the order of $\mu s$) in the combustion chamber. This limitation often gives rise to a range of issues, including inefficient combustion, increased emissions, and unstable flame behavior \cite{LIU2020100636}. The instabilities that arise from poor mixing not only diminish thrust and overall performance but also pose a considerable risk of engine unstart events. It may also result in the formation of localized hotspots that can compromise the structural integrity of the combustor walls, thereby increasing the likelihood of material failure and subsequently limiting the lifespan of the engine. Given the importance of turbulent mixing and combustion processes, a comprehensive understanding of these phenomena is essential for fully realizing the potential of scramjet technology.\\

Remarkable progress has been made over the two past decades in scramjet combustion modeling through the use of Reynolds-Averaged Navier-Stokes (RANS) \cite{xiao_aiaaj_2006,OEVERMANN2000463, pecnik_aiaaj_2012,CHEN2019182} and Large Eddy Simulation (LES) \cite{BERGLUND20072497, YUAN2023108401, YAO2021106941, ZHANG2023128502, shin_iddes_2018} techniques. Early studies employing RANS models focused on evaluating closure strategies to improve predictions of flow and combustion characteristics in scramjets. For instance, the work by Xiao et al \cite{xiao_aiaaj_2006} proposed a new turbulence closure based on variable Schmidt number focusing on modeling supsersonic turbulent mixing in the absence of chemical reactions. This work demonstrated the need for models that can effectively predict turbulent Prandtl and Schmidt numbers as part of the solution. They found that a low Schmidt number could trigger the conditions leading to engine unstart, while a high Schmidt number could cause flame extinction, illustrating the balance needed for combustion analysis. To incorporate turbulence effects and chemical reactions, various combustion models have been developed for supersonic flows, including the presumed Probability Density Function (PDF) model \cite{GONZALEZ_pecs_2017}, the flamelet model \cite{OEVERMANN2000463, pecnik_aiaaj_2012}, and Finite Rate Chemistry (FRC) \cite{DeBoskey2025} approaches, among others. Oevermann et al, in \cite{OEVERMANN2000463}, was among the first to implement the flamelet model in a two-dimensional scramjet combustor simulation, deriving species mass fractions from the flamelet library while solving the mean temperature with the energy equation, enabling its application in complex supersonic flows. Pecnik et al. \cite{pecnik_aiaaj_2012} extended this approach within a RANS framework to study hydrogen jet–crossflow interactions in the HyShot II engine, and Chen et al. \cite{CHEN2019182} further applied it to pulsed fuel injection, showing enhanced mixing tied to injection frequency. More recently, RANS studies using FRC \cite{DeBoskey2025} in high-speed combustion have examined the influence of detailed kinetics and closure strategies on flame structure and combustor efficiency, demonstrating that simplified approaches capture system-level trends but that accurate species composition requires higher-fidelity chemistry. Collectively, these studies underscore advances in RANS-based combustion modeling for scramjets, providing practical tools for system-level design and performance assessment.\\

With advancements in computational power and algorithms, LES has gained traction as a powerful tool for accurately capturing the unsteady dynamics inherent in high-speed turbulent combustion. In LES turbulent scales larger than the grid spacing are resolved, while subgrid scales, which are considered more universal, are modeled along with their impact on the larger scales. This approach enables the capturing of unsteady phenomena such as turbulence–chemistry interactions, shock–shock interactions, shock–boundary-layer interactions, and shock–flame interactions. Consequently, LES is considered a more suitable method for simulating unsteady combustion processes, such as those occurring within ramjet and scramjet engines. The application of LES in capturing the complexities of unsteady combustion processes has been further demonstrated in recent studies \cite{BERGLUND20072497, YUAN2023108401}. In \cite{BERGLUND20072497}, Berglund et al. conducted a comprehensive LES study to investigate supersonic mixing and combustion in a model scramjet combustor, similar to the laboratory scramjet at the Institute for Chemical Propulsion of the German Aerospace Center (DLR), which incorporates a wedge-shaped flameholder for hydrogen fuel injection. The investigation utilizes two flamelet models of varying fidelity to assess the mixing and combustion phenomena occurring downstream of the wedge, demonstrating a strong correlation between the  predictions and experimental findings. Similarly, in \cite{YUAN2023108401}, Yuan et al. explored scramjet engines to better understand flow characteristics, flame stabilization mechanisms, and flame dynamics in separated flows. They utilize LES to analyze a hydrogen-fueled direct-injection scheme for dual-mode scramjet combustion across varying equivalence ratios, emphasizing the critical role of LES in capturing transient phenomena such as flame flashback and oscillation modes during mode transitions. Recently, numerous researchers \cite{YAO2021106941, ZHANG2023128502, shin_iddes_2018} have begun exploring hybrid methods like the Improved Delayed Detached Eddy Simulation (IDDES), which effectively integrate RANS and LES techniques across different flow regions. For example, in \cite{YAO2021106941}, Yao et al. \cite{YAO2021106941} employed IDDES with a zone-based flamelet model (DZFM), demonstrating effective coupling of internal and external flows in high-Mach scramjets. Using IDDES with DZFM, simulations of the DLR combustor revealed significant variations in flame modes and turbulence–chemistry interactions, highlighting the limitations of a single flamelet representation and the need for dynamic flow partitioning and low-Reynolds adjustments. Further, Baknhne et al. \cite{Baknhne2023} applied IDDES with finite-rate chemistry (FRC) and adaptive implicit schemes to hydrogen jets in high-speed transverse vitiated flows. Validation against ONERA’s LAERTE experiments showed that IDDES with FRC accurately captures wall pressure distributions, chemiluminescence, and the sensitivity of combustion dynamics to wall roughness and temperature. While FRC provides the highest fidelity for turbulent combustion, its computational cost remains high, so subgrid-scale (SGS) closures are still needed in practical LES and IDDES applications. These studies demonstrate the growing maturity of IDDES as a high-fidelity tool for scramjet combustion modeling, bridging detailed physics with system-level predictions.\\

Hence, the objective of this work is to investigate the ability of the IDDES model with FRC to accurately predict scramjet propulsion flow path dynamics, including turbulent mixing and combustion. This study employs the US3D solver to model a radical-farming type scramjet based on experiments conducted at the University of Queensland’s T4 wind tunnel \cite{Lorrain_aiaa_2012}, corresponding to a flight Mach number of 10. A novel integrated modeling strategy couples all flow-path components—inlet, fuel injectors, combustor, and nozzle—to capture unsteady chemically reacting flows. Reacting simulations using IDDES with low-dissipation methods are conducted across a range of equivalence ratios. Combustion regimes and fuel–air mixing efficiency are evaluated using the Takeno flame index and a Mixedness metric. This framework enables detailed comparison with experimental observations of combustion patterns and hot-pocket locations, demonstrating its relevance to future hypersonic air-breathing propulsion systems. The paper is organized as follows: Section 2 \ref{sec:intro} introduces an overview of the research topic and a literature review, Section 3 \ref{sec:methods} presents the problem formulation including the flow configuration, governing equations, and numerical methods, Section 4 \ref{sec:results} the results and major findings, and Section 5 \ref{sec:conclusions} presents the conclusions and future work.

\section{Methods}
\label{sec:methods}

\subsection{Experimental and computational setup}
This work uses a scramjet facility in the T4 shock tunnel at the University of Queensland as the reference geometry. The engine studied is an inlet-fueled radical-farming scramjet, previously examined in the literature \cite{Boyce_sw_2012, moura_aiaaj_2020}. In this design, gaseous hydrogen is injected at the inlet to premix with air before entering the combustor. Ideally, mixing occurs upstream, allowing the combustor to focus on ignition and heat release. Ignition-critical radicals form in high-temperature, high-pressure zones—termed hot pockets (HP)—created by shock and expansion wave interactions. These radicals are convected downstream, enabling staged ignition across successive HP regions. Experimental data and further details on radical-farming scramjet combustion in shock tunnels are available in Lorraine et al. \cite{Lorrain_aiaa_2012}.\\

\begin{figure}[hbt!]
    \centering
    \includegraphics[width=0.85\linewidth]{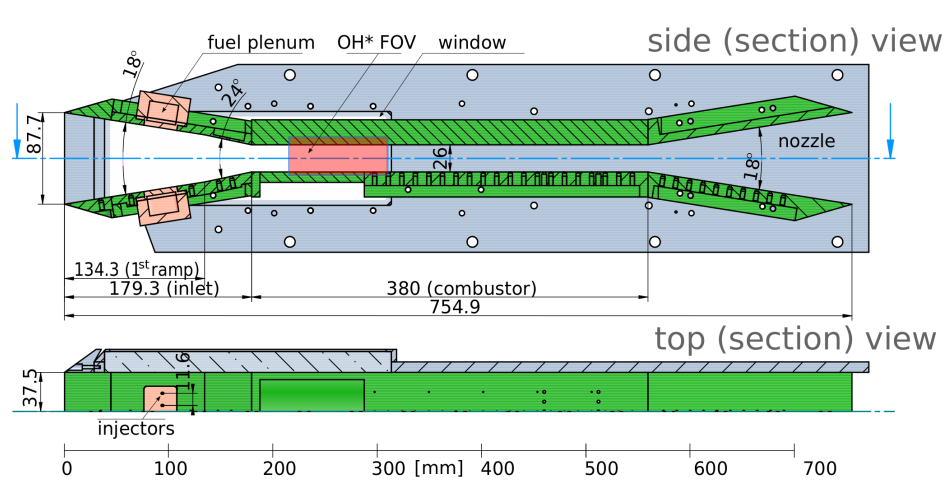}
    \caption{Schematic of the radical farming type scramjet configuration mounted on the T4 wind tunnel at the University of Queensland \cite{Lorrain_aiaa_2012}}
    \label{label:UQscramjet}
\end{figure}

Figure \ref{label:UQscramjet} shows the experimental setup: a two-ramp inlet with 9° and 3° turning angles, a constant-area combustor (26 mm high, 380 mm long), and a 198 mm long single-expansion ramp nozzle angled 9° from horizontal. The inlet ramps measure 136 mm and 46 mm. The internal width is 75 mm, and all sections meet at sharp corners with symmetry about the centerline. Hydrogen was injected sonically via eight cylindrical injectors, located 94 mm downstream of the inlet. Fuel was supplied at 300 K from plenums. The first injector is 20.1 mm from the side wall, with 11.6 mm spacing between ports. Injection occurred at 45° to the local flow, creating elliptical orifices on the ramp surface. Optical access is provided by large side windows, allowing flow visualization up to 120 mm into the combustor. Nitrogen was used for non-reacting tests; air for combustion tests. High-speed fuel-oxidizer interaction led to turbulent mixing. Centerline pressure probes were used to determine wall pressure coefficients \( C_p \), and flow visualization in the optical access region was performed using schlieren and chemiluminescence diagnostics.\\

The experimental facility was utilized to conduct a series of shock tunnel experiments at a total flow enthalpy of 4.3 \, \text{MJ/kg} simulating Mach 9.7 flight at an effective freestream dynamic pressure and altitude of 57.7 \, \text{kPa} and 31.9 \, \text{km}, respectively. Inflow boundary conditions identical to those recorded during the experiment were used at the inlet and are listed in \cite{Lorrain_aiaa_2012} and given as: $p_\infty = 4.1\,\text{kPa}, \; T_\infty = 370\,\text{K}, \; u_\infty = 2830\,\text{m/s}, \; M_\infty = 7.3, \; \dot{m}_{O_2} = 167.3\,\text{g/s}, \; \dot{m}_{H_2} = 16.8\,\text{g/s}$ for equivalence ratio of $\phi = 0.8$.\\

Lastly, other fuel injector inflow conditions were specified to obtain the desired equivalence ratios. These are listed in Table \ref{tab:H2_properties} below, 

\begin{table}[h!]
\centering
\begin{tabular}{lccccc}
\hline
Variable & $\phi = 0.5$ & $\phi = 0.6$ & $\phi = 0.7$ & $\phi = 0.8$ & $\phi = 0.9$ \\
\hline
$\rho_{H_2}$ (kg/m$^3$) & 0.539 & 0.647 & 0.755 & 0.863 & 0.970 \\
$\dot{m}_{H_2}$ (g/s)   & 10.53 & 12.64 & 14.75 & 16.87 & 18.96 \\
$T_{H_2}$ (K)           & 404   & 404   & 404   & 404   & 404 \\
$|\mathbf{u}_{H_2}|$ (m/s) & 777 & 777 & 777 & 777 & 777 \\
\hline
\end{tabular}
\caption{Fuel injector flow conditions for the range of simulated equivalence ratios.}
\label{tab:H2_properties}
\end{table}

\subsection{Numerical Method}
The US3D solver is a highly parallelizable code that employs a finite volume method to solve the governing equations of compressible fluid dynamics for multi-species flows while incorporating finite-rate chemistry. For scale-resolving simulations, minimizing numerical dissipation is essential. Accordingly, low-dissipation stabilized fourth-order inviscid fluxes and centered second-order viscous fluxes are employed. The hybridized flux scheme blends a Kinetic-Energy Consistent (KEC) centered flux with the dissipative component of a shock-capturing scheme to maintain accuracy in both smooth and high-gradient regions. For time integration, a second-order coupled backward difference (BDF) scheme with limited Newton sub-iterations is applied for exceptionally stiff cases. Species diffusion velocities are computed using a constant Lewis number approximation, and viscosity is modeled with Blottner fits combined with the Wilke mixing rule. Turbulence–chemistry interactions are not explicitly considered; chemical production terms are evaluated using filtered-mean data. Further details can be found in \cite{SUBBAREDDY20091347, Candler2015}. 

\subsubsection{Governing Equations}
The compressible forms of the conservation equations for mass, momentum, energy, and species transport are written as:

\begin{equation}
\frac{\partial \bar{\rho}}{\partial t} + \frac{\partial \left( \bar{\rho} \tilde{u}_i \right)}{\partial x_i} = 0,
\end{equation}

\begin{equation}
\frac{\partial \left( \bar{\rho} \tilde{u}_i \right)}{\partial t} + \frac{\partial \left( \bar{\rho} \tilde{u}_i \tilde{u}_j \right)}{\partial x_j} 
= - \frac{\partial \bar{p}}{\partial x_i} + \frac{\partial \tau_{ij}}{\partial x_j} + \frac{\partial \bar{\tau}_{ij}'}{\partial x_j} + \bar{\rho} \tilde{f}_i,
\end{equation}

\begin{equation}
\frac{\partial \left( \bar{\rho} \tilde{e}_t \right)}{\partial t} + \frac{\partial \left( \bar{\rho} \tilde{u}_i \tilde{e}_t \right)}{\partial x_i} 
= \frac{\partial}{\partial x_i} \left( \kappa \frac{\partial \tilde{T}}{\partial x_i} + \bar{q}_i' \right) + \bar{\rho} \tilde{f}_i \tilde{u}_i + \Phi,
\end{equation}

\begin{equation}
\frac{\partial \left( \bar{\rho} \tilde{Y}_k \right)}{\partial t} + \frac{\partial \left( \bar{\rho} \tilde{u}_i \tilde{Y}_k \right)}{\partial x_i} 
= \frac{\partial}{\partial x_i} \left( \bar{\rho} D_k \frac{\partial \tilde{Y}_k}{\partial x_i} + \bar{J}_{i,k}' \right) + \tilde{\dot{\omega}}_k.
\end{equation}

Here, \(\tilde{e}_t = \tilde{e} + \tfrac{1}{2} \tilde{u}_i \tilde{u}_i\) is the Favre-averaged total specific energy, \(\tilde{f}_i\) is the body force per unit mass, \(\kappa\) is the thermal conductivity, and \(\bar{q}_i'\) is the turbulent heat flux. \(\Phi\) denotes viscous dissipation. \(\tilde{Y}_k\) is the Favre-averaged mass fraction of species \(k\), \(D_k\) is the species diffusivity, \(\bar{J}_{i,k}'\) is the turbulent mass flux, and \(\tilde{\dot{\omega}}_k\) is the Favre-averaged chemical source term.\\

The molecular viscous stress tensor is
\begin{equation}
\tau_{ij} = \mu \left( \frac{\partial \tilde{u}_i}{\partial x_j} + \frac{\partial \tilde{u}_j}{\partial x_i} \right) - \tfrac{2}{3} \mu \, \delta_{ij} \frac{\partial \tilde{u}_k}{\partial x_k},
\end{equation}
where \(\mu\) is the dynamic viscosity and \(\delta_{ij}\) is the Kronecker delta.\\

The turbulent Reynolds stresses and turbulent fluxes (Favre-consistent) are expressed as
\begin{equation}
\bar{\tau}_{ij}' = - \bar{\rho} \widetilde{u_i'' u_j''}, \quad
\bar{q}_i' = - \bar{\rho} \widetilde{u_i'' h''}, \quad
\bar{J}_{i,k}' = - \bar{\rho} \widetilde{u_i'' Y_k''},
\end{equation}
where the Favre fluctuations are defined as \(u_i'' = u_i - \tilde{u}_i\), \(h'' = h - \tilde{h}\), and \(Y_k'' = Y_k - \tilde{Y}_k\).\\

The instantaneous molar production rate of species \(i\) (per unit volume) is
\begin{equation}
\dot{\omega}_i = M_i \sum_{j=1}^{N_R} \left( \nu_{ij}'' - \nu_{ij}' \right)
\left( k_{j,f} \prod_{s=1}^{N_s} C_s^{\,\nu_{sj}'} - k_{j,r} \prod_{s=1}^{N_s} C_s^{\,\nu_{sj}''} \right),
\end{equation}
where \(M_i\) is the molar mass of species \(i\), \(N_R\) is the number of reactions, \(N_s\) is the number of species, \(\nu_{sj}'\) and \(\nu_{sj}''\) are the stoichiometric coefficients of species \(s\) in reaction \(j\) (reactants and products, respectively), \(k_{j,f}\) and \(k_{j,r}\) are the forward and reverse rate constants, and \(C_s\) is the molar concentration of species \(s\).\\

For hydrogen combustion, a 12-species, 27-reaction mechanism \cite{Burke2011} is employed to accurately capture the complex chemical kinetics involved. The reaction rate constants for each elementary reaction are expressed in Arrhenius form:
\begin{equation}
k = A \, T^n \, \exp\left(-\frac{E_a}{R T}\right),
\end{equation}
where \(A\) is the pre-exponential factor, \(n\) is the temperature exponent, \(E_a\) is the activation energy, \(R\) is the universal gas constant, and \(T\) is the temperature. This mechanism includes all of the fundamental chain-branching and dissociation reactions, as well as the formation and consumption of intermediate radicals such as HO\(_2\) and H\(_2\)O\(_2\), which are important in controlling ignition and flame propagation. Table \ref{kinetics} summarizes the reactions and their Arrhenius parameters. Previous studies have demonstrated that this mechanism reproduces experimental combustion characteristics of hydrogen-oxygen systems across a wide range of high-pressure conditions \cite{DIEPSTRATEN202422, STROHLE2007125}, making it suitable for computational analysis of hydrogen combustion.\\

\begin{table}[htbp]
\centering
\caption{H$_2$/O$_2$ reaction mechanism including 12 species, 27 reactions, and Arrhenius parameters \cite{Burke2011}}
\label{kinetics}
\begin{tabular}{c l l c c c}
\toprule
\textbf{R\#} & \textbf{Type} & \textbf{Reaction} & \textbf{A (cm$^3$/mol/s)} & \textbf{n} & \textbf{$E_a$ (cal/mol)} \\
\midrule
\multicolumn{6}{l}{\textbf{H$_2$-O$_2$ Chain Reactions}} \\
R1  & Chain & H + O$_2$ $\rightleftharpoons$ O + OH & $1.04\times 10^{14}$ & 0.0 & 15286 \\
R2  & Chain & O + H$_2$ $\rightleftharpoons$ H + OH & $3.82\times 10^{12}$ & 0.0 & 7948 \\
R3  & Chain & O + H$_2$ $\rightleftharpoons$ H + OH & $8.79\times 10^{14}$ & 0.0 & 19170 \\
R4  & Chain & H$_2$ + OH $\rightleftharpoons$ H$_2$O + H & $2.16\times 10^{8}$ & 1.51 & 3430 \\
R5  & Chain & OH + OH $\rightleftharpoons$ O + H$_2$O & $3.34\times 10^{4}$ & 2.42 & -1930 \\
\midrule
\multicolumn{6}{l}{\textbf{H$_2$-O$_2$ Dissociation Reactions}} \\
R6  & Dissoc & H$_2$ + M $\rightleftharpoons$ H + H + M & $4.58\times 10^{19}$ & -1.4 & 104380 \\
R7  & Dissoc & H$_2$ + Ar $\rightleftharpoons$ H + H + Ar & $5.84\times 10^{18}$ & -1.1 & 104380 \\
R8  & Dissoc & H$_2$ + He $\rightleftharpoons$ H + H + He & $5.84\times 10^{18}$ & -1.1 & 104380 \\
R9  & Dissoc & O + O + M $\rightleftharpoons$ O$_2$ + M & $6.17\times 10^{15}$ & -0.5 & 0 \\
R10 & Dissoc & O + O + Ar $\rightleftharpoons$ O$_2$ + Ar & $1.89\times 10^{13}$ & 0.0 & -1788 \\
R11 & Dissoc & O + O + He $\rightleftharpoons$ O$_2$ + He & $1.89\times 10^{13}$ & 0.0 & -1788 \\
R12 & Dissoc & O + H + M $\rightleftharpoons$ OH + M & $4.71\times 10^{18}$ & -1.0 & 0 \\
R13 & Dissoc & H$_2$O + M $\rightleftharpoons$ H + OH + M & $6.06\times 10^{27}$ & -3.322 & 120790 \\
R14 & Dissoc & H$_2$O + H$_2$O $\rightleftharpoons$ H + OH + H$_2$O & $1.01\times 10^{26}$ & -2.44 & 120180 \\
\midrule
\multicolumn{6}{l}{\textbf{Formation and Consumption of HO$_2$}} \\
R15 & Falloff & H + O$_2$ (+ M) $\rightleftharpoons$ HO$_2$ (+ M) & $4.65\times 10^{12}$ & 0.44 & 0 \\
R16 & Cons & HO$_2$ + H $\rightleftharpoons$ H$_2$ + O$_2$ & $2.75\times 10^{6}$ & 2.09 & -1451 \\
R17 & Cons & HO$_2$ + H $\rightleftharpoons$ OH + OH & $7.08\times 10^{13}$ & 0.0 & 295 \\
R18 & Cons & HO$_2$ + O $\rightleftharpoons$ O$_2$ + OH & $2.85\times 10^{10}$ & 1.0 & -724 \\
R19 & Cons & HO$_2$ + OH $\rightleftharpoons$ H$_2$O + O$_2$ & $2.89\times 10^{13}$ & 0.0 & -497 \\
\midrule
\multicolumn{6}{l}{\textbf{Formation and Consumption of H$_2$O$_2$}} \\
R20 & Form & HO$_2$ + HO$_2$ $\rightleftharpoons$ H$_2$O$_2$ + O$_2$ & $4.20\times 10^{14}$ & 0.0 & 11982 \\
R21 & Form & HO$_2$ + HO$_2$ $\rightleftharpoons$ H$_2$O$_2$ + O$_2$ & $1.30\times 10^{11}$ & 0.0 & -1629 \\
R22 & Falloff & H$_2$O$_2$ (+ M) $\rightleftharpoons$ OH + OH (+ M) & $2.00\times 10^{12}$ & 0.9 & 48749 \\
R23 & Cons & H$_2$O$_2$ + H $\rightleftharpoons$ H$_2$O + OH & $2.41\times 10^{13}$ & 0.0 & 3970 \\
R24 & Cons & H$_2$O$_2$ + H $\rightleftharpoons$ HO$_2$ + H$_2$ & $4.82\times 10^{13}$ & 0.0 & 7950 \\
R25 & Cons & H$_2$O$_2$ + O $\rightleftharpoons$ OH + HO$_2$ & $9.55\times 10^{6}$ & 2.0 & 3970 \\
R26 & Cons & H$_2$O$_2$ + OH $\rightleftharpoons$ HO$_2$ + H$_2$O & $1.74\times 10^{12}$ & 0.0 & 318 \\
R27 & Cons & H$_2$O$_2$ + OH $\rightleftharpoons$ HO$_2$ + H$_2$O & $7.59\times 10^{13}$ & 0.0 & 7270 \\
\bottomrule
\end{tabular}
\end{table}

Turbulence is modeled using the Improved Delayed Detached Eddy Simulation (IDDES), a hybrid RANS–LES method \cite{SHUR20081638}. Near walls, it reduces to the Spalart–Allmaras Catris RANS model \cite{CATRIS20001}, capturing mean flow efficiently. In regions of intense turbulence, it transitions to LES, resolving large-scale structures and modeling subgrid stresses with the Vreman model \cite{VREMAN2004}. The RANS–LES switch is governed by the hybrid length scale
\begin{equation}
l_{IDDES} = l_{RANS} - f_d \, \max\big(0, l_{RANS} - C_{DES} \, \Delta \big),
\end{equation}
where \(l_{RANS}\) is the RANS length scale, \(\Delta\) the local grid spacing, \(C_{DES}\) a model constant, and \(f_d\) a delay function that suppresses premature LES near walls. This approach confines LES to appropriate flow regions while maintaining RANS behavior near boundaries. Combined with detailed chemistry, IDDES enables accurate simulation of unsteady turbulent hydrogen combustion.

\subsection{Grid convergence analysis}
The computational configuration employs a three-dimensional model derived from the schematic of the experimental facility, see Figure \ref{label:UQscramjet}. This model is constrained to a \( \frac{1}{4} \) representation of the scramjet, incorporating symmetry boundaries to optimize computational efficiency. Dirichlet boundary conditions are applied at the inlet, encompassing both the inflow and the H$_2$ fuel jets. At the outlet, all variables are extrapolated from the interior cells, while a no-slip, isothermal 300K condition is enforced on all walls.\\

\begin{figure}[hbt!]
    \centering
    \includegraphics[width=0.7\linewidth]{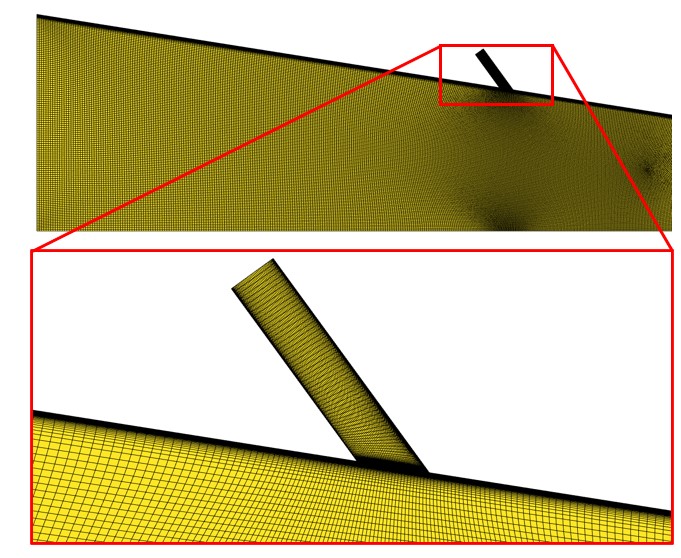}
    \caption{Visualization of the structured hexahedral grid highlighting the fuel injector and near-wall region. Top: view of ramp-injector region mesh. Bottom: view of injector mesh cross section.}
    \label{fig:meshviz}
\end{figure}

The three-dimensional hexahedral grids used in this work were made using the GoHypersonic Inc. LINK3D meshing software \cite{link3d_meshing}, which enables efficient, parallel structured grid generation that is amenable to geometric modifications and design optimization. The baseline grid system considered consisted of $31$ million cells, with wall spacing yielding $y^{+}$ values well below one to ensure adequate resolution - see Figure \ref{fig:meshviz} for a visualization of our grid details with emphasis in the fuel injector and near-wall region. Several different grid systems were evaluated, each with successively more cells than the last, in order to determine the minimum number of points necessary to resolve the relevant flow features. This grid-convergence analysis was conducted based on the experimental case with an equivalence ratio of 0.8. The analysis compared the baseline mesh with $31$ million cells (coarse) and two refined meshes with $45$ million cells (medium) and $62$ million cells (fine). Both grid systems maintained the same wall-normal spacing with $y^{+}$ < 1 across the entire surface, but the fine resolution case had approximately 5 times more cells in the streamwise and spanwise directions. The variation percentages on the coefficient of surface pressure for each grid systems indicated that further resolution refinements did not impact the simulation outcomes.\\ 

\begin{figure}[hbt!]
    \centering
    \begin{minipage}{0.5\linewidth} 
        \centering
        \includegraphics[width=\linewidth]{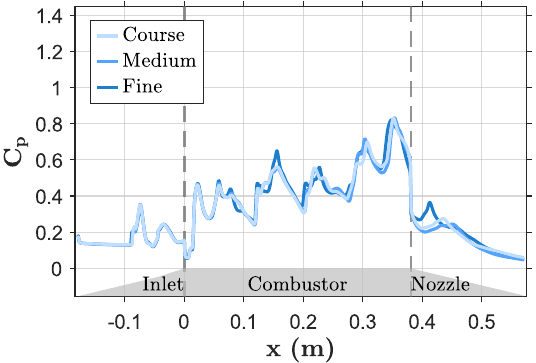}
        \caption*{} 
    \end{minipage}%
    \hfill 
    \begin{minipage}{0.5\linewidth} 
        \centering
        \includegraphics[width=\linewidth]{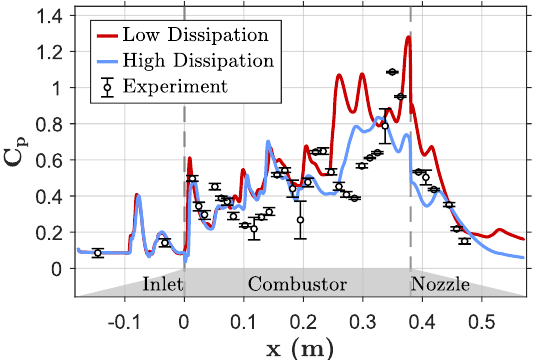}
        \caption*{} 
    \end{minipage}
    \caption{Numerical characteristics of flow solution. (Left) Grid convergence analysis for a coarse $31$ Mil., medium $45$ Mil., and fine $62$ Mil. cells for $C_{p}$ coefficient of pressure along with scramjet engine. (Right) Effect of low-dissipation numerics on solution accuracy.}
    \label{fig:meshconv}
\end{figure}

Figure \ref{fig:meshconv} (left) presents the variations of the centerline pressure coefficient along the surface for both the coarse, medium and refined grids. The curves for both grids closely overlap, which confirms that the resolution of fine mesh is sufficiently refined for the objectives of this study. Specifically, the $C_{p}$ coefficient exhibits mosts changes of less than $1-3\%$ when the mesh resolution was increased. This minimal variation suggests that the results are well-converged with respect to grid refinement.Furthermore, the overall impact of additional grid resolution on thermal and reacting flow paramters —such as transport of chemical species and heat flux were negligible. These findings indicate that further refinement of the grid would not lead to significant changes, nor would it alter the combustion characteristics in a meaningful way. Therefore, we conclude that the fine grid resolution provides an accurate and efficient representation of the flow physics for the purposes of this analysis, and additional refinement would offer diminishing returns.\\

Although the simulation showed good convergence behavior and reasonable comparison with experimental results, a slight under-prediction of the peak pressure rise in the combustor was observed. The discrepancies in the pressure rise are shown in Figure \ref{fig:meshconv} (right). To address this, we incorporated low-dissipation numerical methods, including the Ducros sensor \cite{DUCROS1999517} and the second-order accurate backward differntiation formula (BDF2) time-accurate scheme, to reduce numerical errors.  This adjustment, with an error margin of less than 2\%, ensures that the flow is better captured (as compared to low dissipation method), by capturing more of the steep gradients and leading to more physically realistic simulation results. Hence, the remaining of the simulations are conducted using the low-dissipation methods. \\

\newpage
\clearpage
\section{RESULTS} 
\label{sec:results}
This section presents results from US3D simulations utilizing IDDES scale-resolving turbulence modeling, coupled with a finite chemistry model for hydrogen fuel. The focus is on key advancements in predicting auto-ignition and characterizing the fuel distribution within the combustor. The analysis specifically investigates flow development and the dynamics of fuel-air mixing, as visualized through cross-sectional planes at various streamwise locations along the combustor. Additionally, comparisons between experimental and computational schlieren visualizations are provided, along with an exploration of the effects of equivalence ratio and flame structure on the overall combustion performance.

\subsection{Hotspot Formation and Combustion Evolution}
\begin{figure}[hbt!]
    \centering
    \includegraphics[width=0.92 \linewidth]{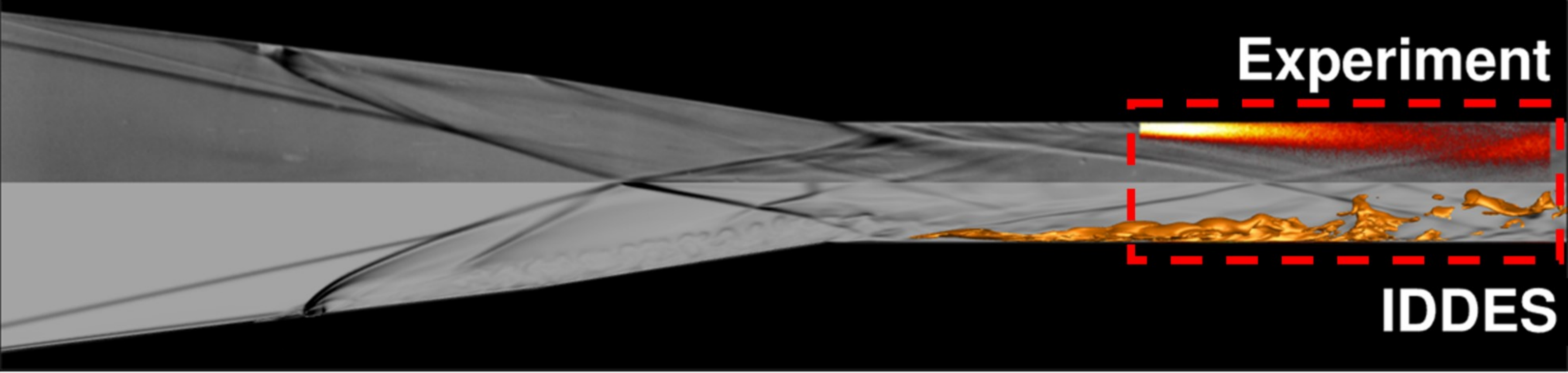}
    \caption{A comparison of experimental Schlieren imaging and OH chemiluminescence measurements with numerically generated Schlieren data at the centerplane, along with the OH mass fraction isosurface at 1\% for $\phi = 0.6$.}
    \label{fig:schlieren}
\end{figure}

Experiments by Lorrain et al. \cite{Lorrain_aiaa_2012} demonstrated that the hotspot behind the first shock reflection within the isolator governs auto-ignition in the scramjet. Our analysis showed that steady-state simulations misplace this hotspot, causing ignition to shift further downstream. By using IDDES, we were able to address this issue. A comparison between the computational schlieren image and the experimental data, shown in Figure \ref{fig:schlieren}, reveals strong agreement in the shock structure, including the shocks generated by the transverse hydrogen jets on the ramp. The image also includes chemiluminescence data from the experiment. By comparing OH concentrations (via an iso-surface of OH mass fraction 1\%) with the chemiluminescence, auto-ignition is observed earlier in the isolator in the simulation. Although the experimental imaging shows the first reaction products further downstream than the simulation, this is due to the limitations of their imaging window. The authors describe auto-ignition as occurring behind the first shock reflection in the isolator, which is precisely where the simulation predicts it to occur.

\begin{figure}[hbt!]
    \centering
    \includegraphics[width=0.75\linewidth]{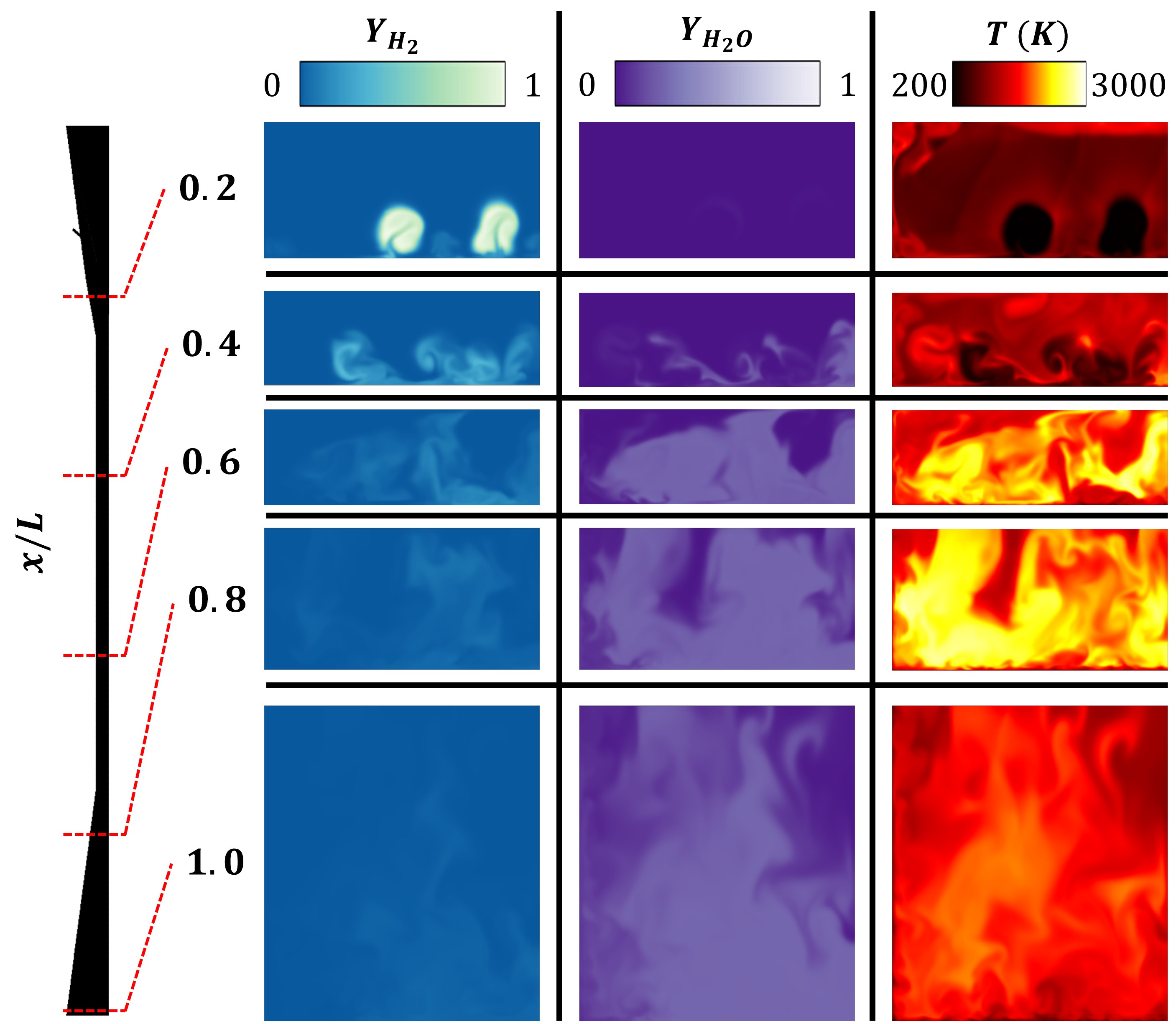}
    \caption{Cross-sectional planes along the scramjet flow path ($x/L = 0.20 - 1.0$) showing hydrogen dispersion and combustion, with lobes fragmenting downstream of injection and becoming more uniformly mixed through the isolator/combustor. at $\phi = 0.8$.}
    \label{fig:crossplane1}
\end{figure}

Figure \ref{fig:crossplane1} illustrates cross-sectional planes along the engine flow path, with the locations of the stations from $x/L = 0.20$ to $x/L = 1.0$ shown in the sketch on the left of the figure. The first row ($x/L = 0.2$) is located downstream of the fuel injectors on the ramp, followed by two rows within the isolator/combustor region ($x/L = 0.4-0.6$), and the final two rows ($x/L = 0.8-1.0$) positioned at the nozzle ramp and the domain exit. These stations provide a detailed view of the evolution of hydrogen dispersion and combustion as the flow progresses along the combustor's length. In the first column of the figure, the hydrogen concentration lobes are seen to break apart on the ramp surface. This fragmentation is evident from the split concentration observed within each lobe, suggesting increased fuel mixing with the surrounding air. By the time the flow reaches the second and third stations, substantial dispersion of the hydrogen fuel is noticeable, particularly within the isolator/combustor region, where the fuel becomes more uniformly distributed across the flow field.\\

In the second column of the figure, the corresponding water vapor concentration is shown, representing a key byproduct of the combustion process. Notably, the presence of water vapor at $x/L = 0.4$ is indicative of combustion occurring upstream of this location, suggesting that auto-ignition has already taken place. This observation is further supported by a rise in the temperature field, as seen in the corresponding temperature maps. As the flow progresses towards the exit plane, the hydrogen fuel is nearly entirely consumed, and the temperature fields start to cool off and disperse relative to the $x/L = 0.8$ station. This suggests that the majority of the combustion reactions are concentrated within the isolator/combustor region, confirming that the combustor is operating efficiently and that the auto-ignition process is active within the targeted zone.

\subsection{Effect of Equivalence Ratio}
\begin{figure}[hbt!]
    \centering
    \includegraphics[width=0.9\linewidth]{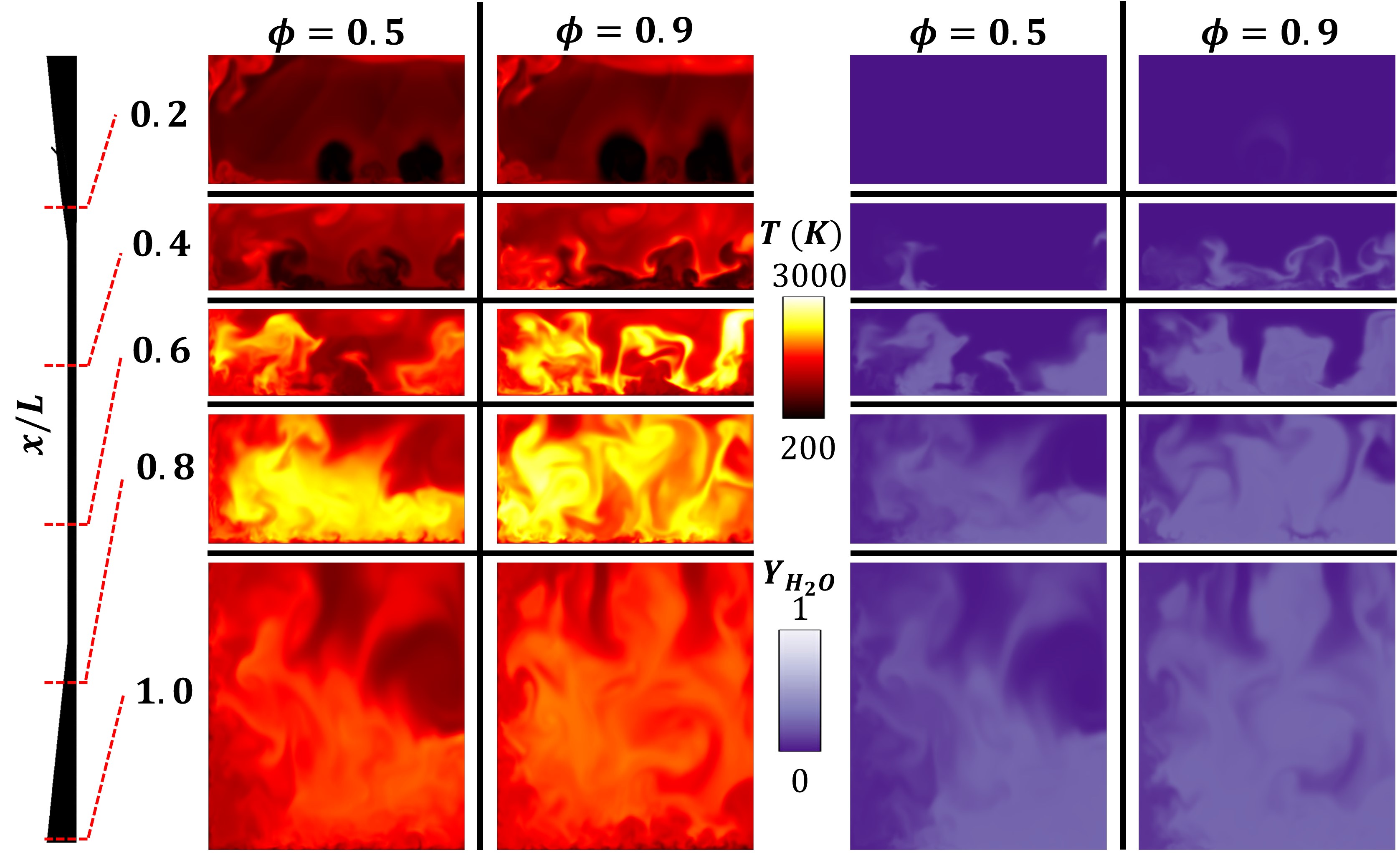}
    \caption{Cross-sectional planes along the scramjet flow path ($x/L = 0.20 - 1.0$) of temperature and water vapor distributions for $\phi=0.5$ and $\phi=0.9$, showing higher temperatures and faster fuel consumption at the richer condition.}
    \label{fig:cutplanes2}
\end{figure}

In Figure \ref{fig:cutplanes2}, the left side of the figure presents two subcolumns, each showing the temperature distributions for equivalence ratios of 0.5 and 0.9, respectively. The right side displays the corresponding mass fraction of water vapor for each equivalence ratio. The temperature contours reveal notable differences between the two conditions. For the equivalence ratio of 0.5, the magnitude of the temperature distribution is relatively lower compared to the higher equivalence ratio of 0.9, where the combustion temperatures are higher. This indicates a more intense combustion process at the higher equivalence ratio, likely because of the increased compressibility and turbulent activity, which results in a higher heat release and thus elevated temperatures. In both cases, the appearance of water vapor at $x/L = 0.4$ signals that auto-ignition has already taken place upstream of this location. The rise in temperature, particularly for the higher equivalence ratio, is further evidence of this. As the flow progresses towards the exit plane, the hydrogen fuel is nearly fully consumed and the values of the temperature distribution begin to decrease, particularly after $x/L = 0.8$. For the equivalence ratio of 0.9, this temperature decline is more pronounced, reflecting the faster consumption of the fuel and the completion of combustion within a shorter distance compared to the lower equivalence ratio.\\

These observations suggest that most combustion reactions occur within the isolator/combustor region, confirming efficient combustor operation for both equivalence ratios. However, the higher combustion temperatures observed at $\phi = 0.9$ indicate a more complete and intense combustion process, while the lower temperatures at $\phi = 0.5$ suggest a less complete combustion, likely due to differences in compressibility and turbulence activity. The results highlight how varying the equivalence ratio affects the combustion temperature and water vapor formation, with the higher equivalence ratio promoting more intense combustion and higher temperatures.\\

\begin{figure}[hbt!]
    \centering
    \begin{minipage}{0.5\textwidth}
        \centering
        \includegraphics[width=\linewidth]{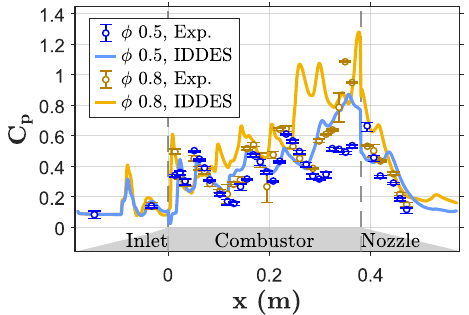}
    \end{minipage}%
    \hfill
    \begin{minipage}{0.5\textwidth}
        \centering
        \includegraphics[width=\linewidth]{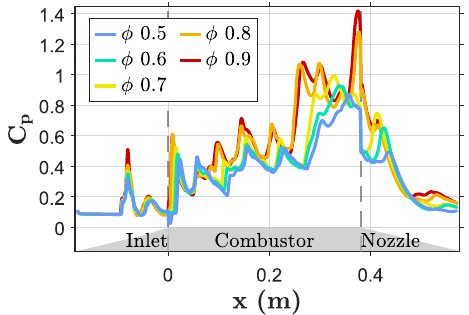}
    \end{minipage}
    \caption{Comparison of IDDES results: (left) validation of $C_p$ distributions for two equivalence ratios; (right) $C_p$ distributions for all equivalence ratios.}
    \label{fig:cp_comparison12}
\end{figure}

Figure \ref{fig:cp_comparison12} (left) compares $C_p$ distributions for $\phi = 0.5, 0.8$ with experiments, while Figure \ref{fig:cp_comparison12} (right) shows simulated distributions across $\phi = 0.5$–0.9. At $\phi = 0.8$, the distribution exhibits pronounced peaks, indicative of stronger variability in the pressure field driven by localized flow instabilities and unsteady combustion dynamics, with simulations reasonably reproducing most of these features. In contrast, at $\phi = 0.5$, the distribution is smoother with reduced variability, and the simulations align more closely with experimental measurements, suggesting lower sensitivity to localized fluctuations at the leaner condition. As $\phi$ increases from 0.5 to 0.9, the simulated distributions develop sharper gradients and enhanced variability, consistent with stronger combustion-induced pressure fluctuations and increased coupling between thermal and fluid-dynamic effects. Overall, the simulations capture the dominant flow behavior across all equivalence ratios, with discrepancies largely attributable to uncertainties in the experimental measurements, the lack of detailed information regarding turbulence levels, and possible flow non-uniformity due to tunnel related effects. 

\newpage
\subsection{Flame structure}
\begin{figure}[hbt!]
    \centering
    \begin{minipage}{0.8\linewidth} 
        \centering
        \includegraphics[width=\linewidth]{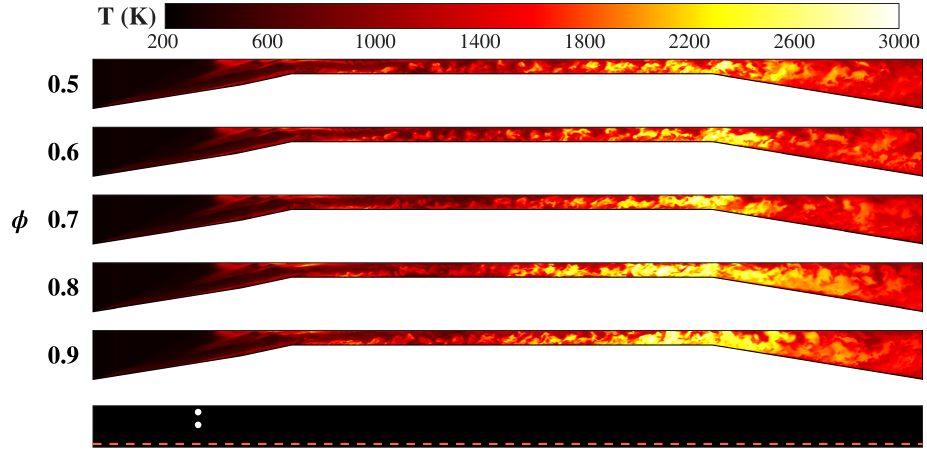}
        \caption*{} 
    \end{minipage}
    \\ 
    \begin{minipage}{0.8\linewidth} 
        \centering
        \includegraphics[width=0.9\linewidth]{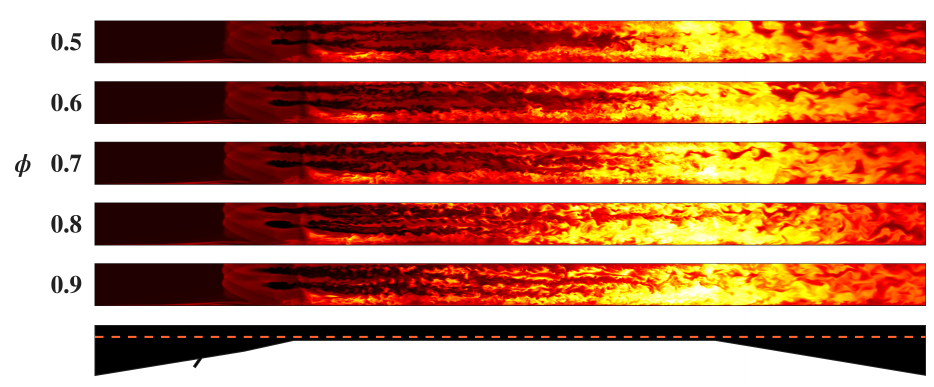}
        \caption*{} 
    \end{minipage}
    \caption{Visualization of effect of equivalence ratio on instantaneous temperature on (top) side view, (bottom) top view 2mm off the wall.}
    \label{fig:inst-temp}
\end{figure}

Figure \ref{fig:inst-temp} presents detailed visualizations of the instantaneous temperature field and flame structure, shown from both side and top views for each equivalence ratio considered. The images reveal spatial variations in flame topology and intensity, highlighting regions of elevated temperature and areas where the flame front is weaker or more diffuse. Comparison of equivalence ratios demonstrates the influence of the fuel-air mixture on flame stabilization, the location of peak heat release, and the overall progression of combustion along the engine flow path. Figure \ref{fig:inst-temp} (top) correspond to a plane located 3 mm from the side wall, selected because it intersects the radical rich region and shows the dependence of the ignition onset location on the equivalence ratio.  Figure \ref{fig:inst-temp} (bottom) shows a top view of temperature contours on horizontal slices along the combustor length, taken 2 mm from the combustor surface. These slices further clarify the effect of equivalence ratio on combustion behavior: at higher equivalence ratios, ignition occurs further upstream, and elevated temperatures persist deeper into the nozzle, indicating stronger combustion effects.\\

Together, these visualizations provide a comprehensive representation of the interplay between flow dynamics, fuel-air mixing, and chemical reaction processes within the scramjet combustor.

\begin{figure}[hbt!]
    \centering
    \includegraphics[width=0.6\textwidth]{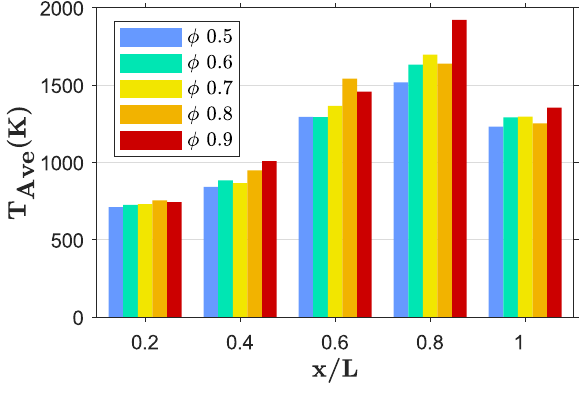}
    \caption{Averaged temperatures for various equivalence ratios at different stations along the scramjet engine}
    \label{fig:cp_temp_comparison}
\end{figure}

Figure \ref{fig:cp_temp_comparison} shows the spatially averaged temperature profiles at each measurement station, highlighting the dependence on equivalence ratio: higher $\phi$ values correspond to higher temperatures. In all cases, the temperature rises steadily through the combustor, reaching a maximum near $x/L = 0.8$, where combustion is nearly complete, and then levels off. As the flow passes into the nozzle, the temperature drops sharply due to gas expansion and the associated cooling. This reduction is more pronounced at higher equivalence ratios, consistent with the greater heat release and more intense combustion upstream.\\

To further examine the combustion phenomenon, a more detailed look at the mechanisms driving the reaction was conducted. This analysis involved calculating the Mixedness and Flame Index as outlined by Takeno \cite{Takeno1996}. The Mixedness parameter $Z_{FO}$ is used to determine how much the unburnt fuel and air have interacted prior to combustion where values less than zero are lean zones and greater than zero are fuel rich zones. They are defined as follows:

\textbf{Mixedness $Z_{FO}$ \cite{Takeno1996}}: 
\begin{equation}
    Z_{FO} = 
    \begin{cases} 
        \frac{Y_O}{j}, & \text{if } \frac{Y_O}{j} \le Y_F \\[1.5mm]
        - Y_F, & \text{if } \frac{Y_O}{j} \ge Y_F
    \end{cases}
\end{equation}
where $Y_F$ and $Y_O$ are the mass fractions of fuel and oxidizer, respectively, and $j$ is a scaling factor defined as $j = \frac{m_O v_O}{m_F v_F}$, where $m_F$ and $m_O$ are the molar masses and $v_F$ and $v_O$ are the stoichiometric coefficients of fuel and oxidizer, respectively.\\

A  metric for analyzing the reactions is an extension of the Mixedness parameter deemed the Takeno Flame Index (TFI). 

\textbf{Takeno flame index $G_{FO}$} \cite{Takeno1996}:
\begin{equation}
    G_{FO} = \nabla Y_F \cdot \nabla Y_O
\end{equation}

This represents the inner product of the gradients of the mass fractions of the fuel (H$_2$) and the oxidizer (O$_2$), quantifying both the alignment of these gradients and their magnitudes. Positive values correspond to premixed flames, where the fuel and oxidizer are already well mixed and convect together with the flow, resulting in aligned gradients. Negative values indicate non-premixed (diffusion-like) combustion, where fuel and oxidizer are initially unmixed and their gradients are opposed at the interface between streams. Values near zero correspond to transitional or mixed combustion regimes. In premixed reactions, turbulent mixing drives the combustion process, whereas in non-premixed reactions, combustion is sustained primarily by diffusion. \\

\begin{figure}[hbt!]
    \centering
    \includegraphics[width=0.9\linewidth]{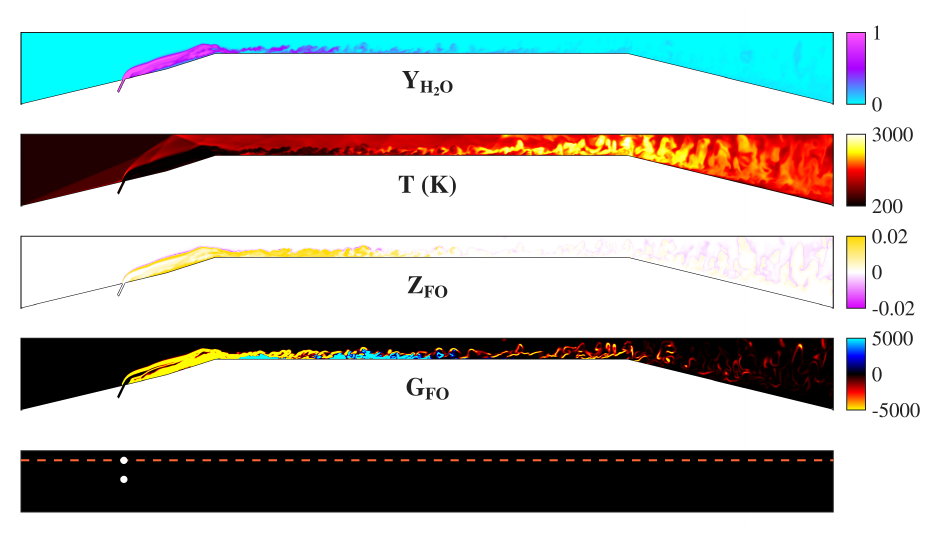}
    \caption{Spatial distributions of the instantaeous hydrogen fuel mass fraction, temperature, mixedness, and Takeno flame index within the scramjet flowpath at $\phi$ = 0.8}
    \label{fig:allmixTFI}
\end{figure}

Figure \ref{fig:allmixTFI} shows the spatial distribution of hydrogen fuel in the scramjet flowpath, along with instantaneous temperature fields, mixedness, and the Takeno flame index at equivalence ratio $\phi$ = 0.8. The fuel distribution illustrates how hydrogen penetrates and spreads within the core flow, the temperature fields indicate zones of combustion and heat release, the mixedness parameter characterizes the local degree of fuel–air mixing, and the Takeno flame index identifies regions of premixed and non-premixed burning. Most of the reaction zone, especially downstream of the isolator entrance, spans multiple combustion regimes. The auto-ignition zone exhibits partially premixed behavior near the wall (at the first hot pocket), surrounded by non-premixed regions, showing enhanced fuel–air mixing before ignition. This emphasizes the need to model transverse fuel injection, as earlier studies \cite{McGuire2008} overlooked this factor in capturing ignition behavior.\\

\begin{figure}[hbt!]
    \centering
    \includegraphics[width=0.9\linewidth]{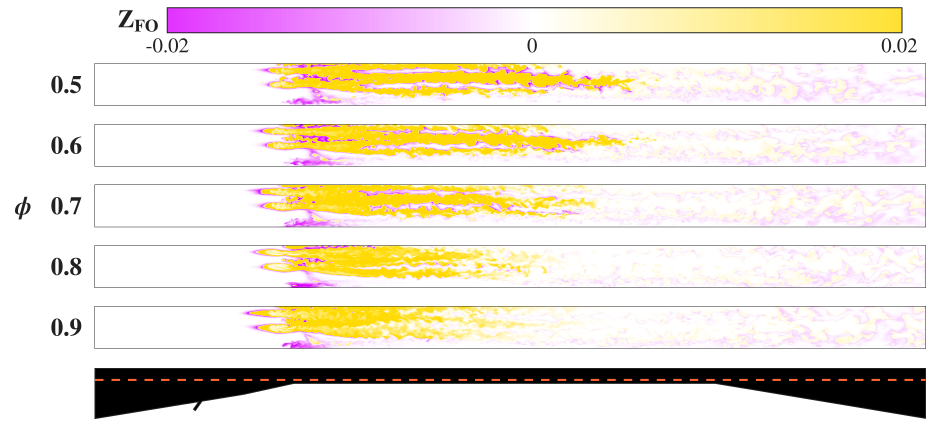}
    \caption{A comparison of the mixedness parameter in the scramjet for equivalence ratio varying from $\phi$ = 0.5 to 0.9 }
    \label{fig:mixrdness}
\end{figure}

Figure~\ref{fig:mixrdness} presents an instantaneous field of \( Z_{FO} \) for equivalence ratios $\phi$ ranging from 0.5 to 0.9. The data show that combustion predominantly occurs under fuel-rich conditions, with heat release concentrated in the cores of the fuel jets rather than along their edges, where the mixture is lean. This localization suggests that mixing is still developing in the near-field, with increasing entrainment of oxidizer as the flow evolves downstream. Toward the aft end of the combustor and into the nozzle, the mixture transitions to fuel-lean as turbulent mixing and geometric expansion promote further dilution and homogenization of the flow.\\

\begin{figure}[hbt!]
    \centering
    \includegraphics[width=0.9\linewidth]{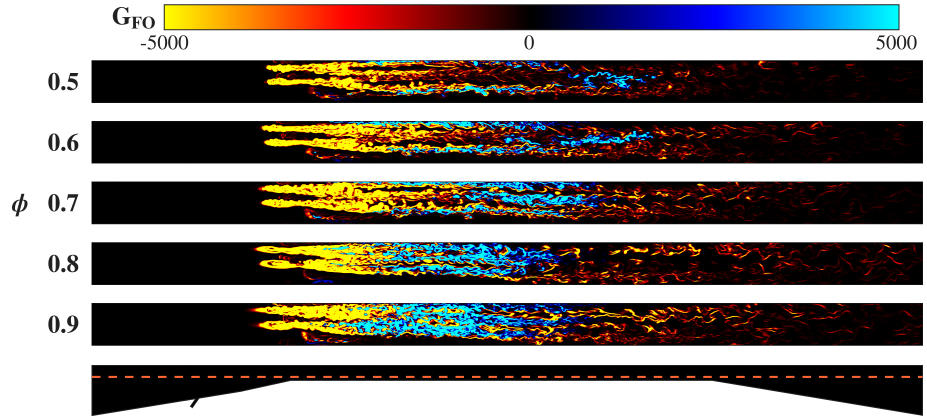}
    \caption{A comparison of Takeno Flame Index in the scramjet for equivalence ratio varying from $\phi$ = 0.5 to 0.9}
    \label{fig:takeno}
\end{figure}

Figure~\ref{fig:takeno} shows contours of the Takeno flame index, used to distinguish premixed from non-premixed combustion. Negative values correspond to non-premixed regions, while positive values indicate premixed combustion, with magnitude increasing as local fuel and oxidizer gradients become more aligned. Three distinct regions are observed: a central diffusion-dominated zone, a premixed core, and an outer layer where diffusion-driven reactions occur adjacent to premixed structures. As the equivalence ratio increases, the premixed region broadens, indicating a growing contribution from premixed combustion. This trend reflects a shift in flame structure, with higher \(\phi\) promoting more extended premixed zones within the turbulent field.

\subsection{Heat-Release Zones and Flame Stabilization Mechanism}
\begin{figure}[hbt!]
    \centering
    \begin{minipage}{0.8\linewidth} 
        \centering
        \includegraphics[width=\linewidth]{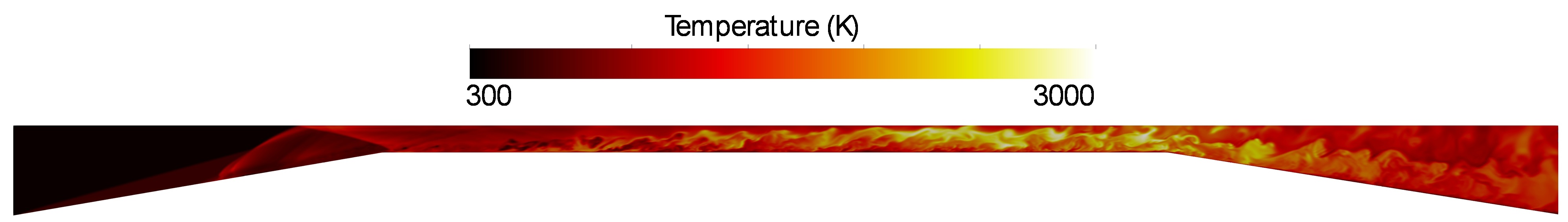}
        \caption*{} 
    \end{minipage}
    
    \vspace{-0.6cm} 
    
    \begin{minipage}{0.8\linewidth} 
        \centering
        \includegraphics[width=\linewidth]{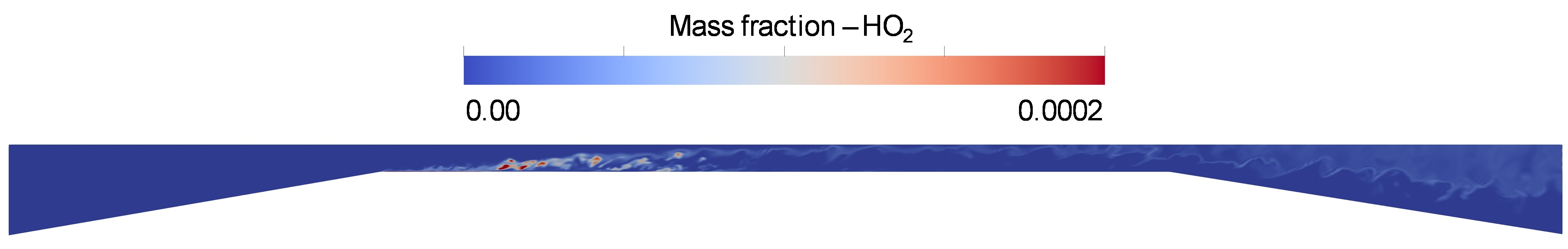}
        \caption*{} 
    \end{minipage}
    \caption{Visualization of instantaneous temperature field and mass fraction of the intermediate species - $HO_2$ leads to an immediate observation that the regions responsible for heat-release do not coincide with regions of highest flame temperature.}
    \label{fig:inst-temp-ho2}
\end{figure}

Given the short combustion time scales and the highly unsteady nature of the flow field, identifying key heat-release zones and determining the flame stabilization mechanism is challenging. Visualization of the instantaneous temperature field and mass fraction of an intermediate species $HO_2$ (hydroperoxyl radical) in Figure \ref{fig:inst-temp-ho2} clearly shows that while the zones of maximum heat release are located immediately downstream of the inlet ramp, pockets of highest temperature are observed towards the end of the combustion chamber. A complete disconnection between these fields poses a challenge for identification of the flame based on such instantaneous contours. As such, it becomes immediately apparent that temperature contour alone is a poor indicator of reaction zones or the flame stabilization mechanism. 

In order to address this challenge, a more rigorous approach to better identify reaction zones is adopted by conducting the Chemical Explosive Mode Analysis (CEMA) \cite{cema}. Chemical explosive modes are associated with positive eigenvalues, $\lambda_{exp}$, of the Jacobian, $J_{\omega}$, of the chemical source term $\omega$, which may be represented as :

\begin{equation}
Re(\lambda_{exp}) > 0
\end{equation}

where the chemical source terms associated with the kinetic mechanism may be assembed as a matrix, which can be represented as follows: 

\begin{equation}
\frac{Dy}{Dt} = \omega(y)
\end{equation}

where y includes the species concentrations. The Jacobian of the matrix $\omega(y)$ is then calculated at all grid points to determine the explosive mode at all spatial locations. Further details of the model may be found here \cite{cema}.

\begin{figure}[hbt!]
    \centering
    \includegraphics[width=0.95\linewidth]{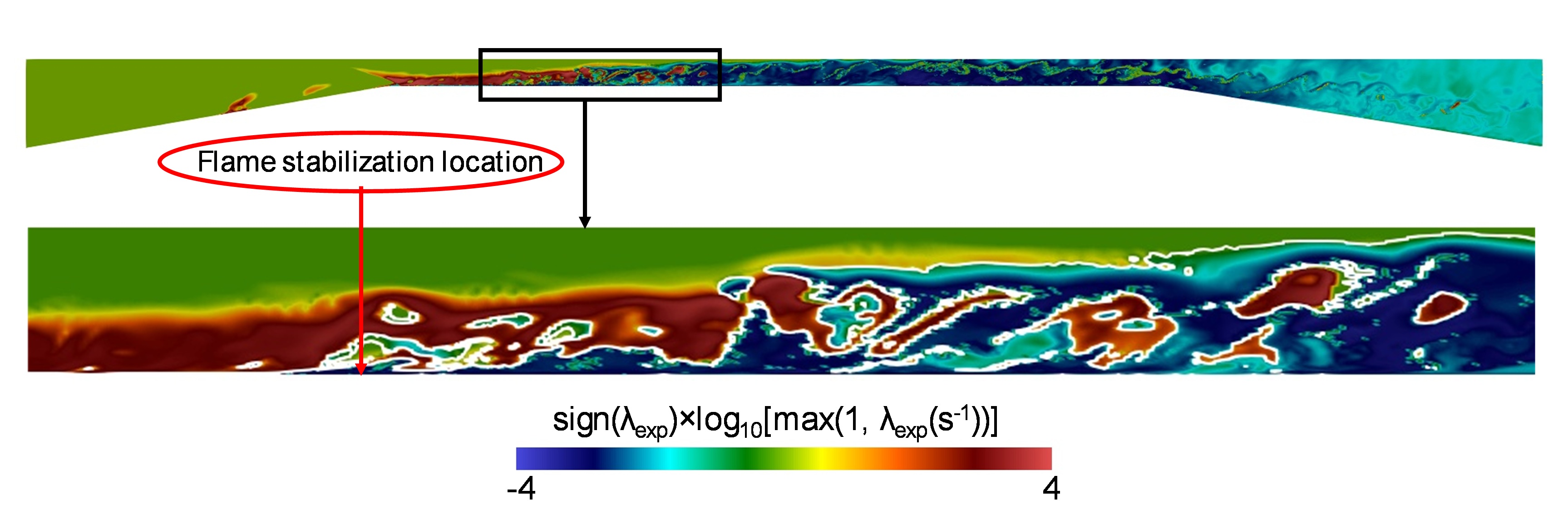}
    \caption{Contour plot of the eigenvalue associated with the explosive mode allows identification of the flame front, as demarkated by the isocontour line corresponding to $\lambda_{exp} = 0$}
    \label{fig:cema}
\end{figure}

In Figure \ref{fig:cema}, visualization of the explosive mode shows that the explosive region is predominantly confined to regions immediately downstream of the ramp, as evidenced by positive values of $\lambda_{exp}$ in log scale. The visualization further shows that further downstream (up to the nozzle region), the flowfield is in a non-explosive regime. Hence, this analysis suggests that most of the combustion occurs immediately downstream of the inlet ramp. These gases then convect downstream, with combustion reaching full completion along the path, thus leading to the presence of high temperature zones much closer to the exit. In fact, a back-of-the-envelope calculation shows that the time scale associated with the eigenvalue of the explosive mode $1/\lambda_{exp}$ is of the same order of magnitude as the convection time scale for flow reaching the combustor exit, which further explains why despite reaction zones existing immediately downstream of the inlet ramp, majority of the high temperature zones are present close to the combustor exit. 

In addition, the white isocontour line corresponding to $\lambda_{exp} = 0$ demarcates the explosive region from the non-explosive region, and as such marks the flame location. Tracking this isoline, it becomes evident that the flame is stabilized close to the combustor wall immediately downstream of the inlet ramp, as marked in figure \ref{fig:cema}. Next, in order to identify the flame stabilization mechanism, a Damkohler number defined based on the local scalar dissipation rate is defined as :

\begin{equation}
Da = \lambda_{exp}.\chi^{-1}
\end{equation}
 
where $\chi$ is defined as $2 \alpha|\nabla \zeta|^2$, where $\alpha$ is the thermal diffusivity and $\zeta$ is the mixture fraction. Such a definition of the Damkohler number compares the time scale of the explosive mode compared with transport.

\begin{figure}[hbt!]
    \centering
    \includegraphics[width=0.95\linewidth]{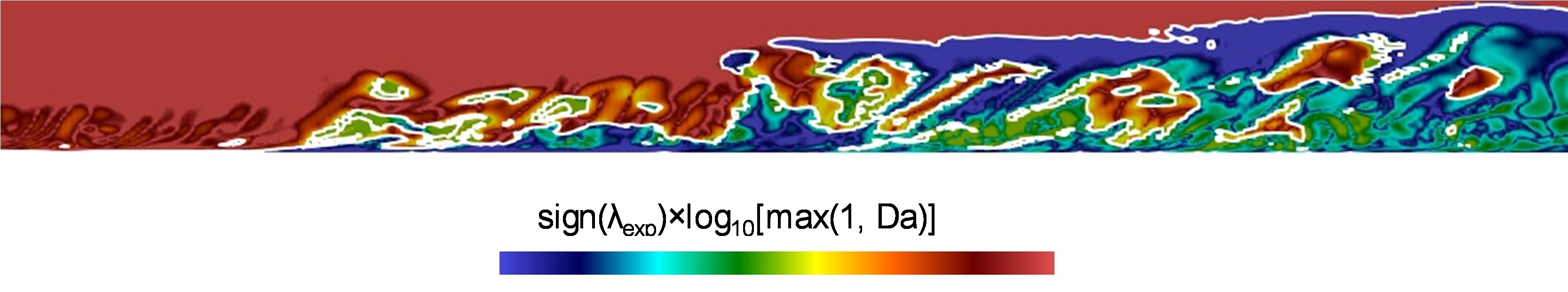}
    \caption{Iso-contour line corresponding to the explosive mode $\lambda_{exp} = 0$ overlaid on top of spatial distribution of the Damkohler number}
    \label{fig:cema_vs_Da}
\end{figure}

Figure \ref{fig:cema_vs_Da} shows the $\lambda_{exp} = 0$ isocontour from CEMA overlaid on the logarithmic contour of $Da$. The close alignment of the explosive mode boundary with regions where $Da \gg 1$ highlights that stabilization occurs through autoignition driven by rapid chemical kinetics, rather than by flame-front propagation.

Since turbulence–chemistry interactions are not modeled in the present study, the influence of turbulence on flame characteristics is neglected. While this omission is expected to affect the mean chemical reaction rates and flame thickness to some extent, the thin boundary layer associated with the high inflow Reynolds number of approximately 220,000 allows the core of the inlet flow to be treated, to some degree, as laminar and inviscid. Turbulence–chemistry interactions notwithstanding, their effect on the CEM analysis is limited to the accuracy with which the species field is predicted by the present chemistry solver, and as such plays no direct role in the CEM analysis, which considers only the sensitivity of the chemical kinetics to changes in species concentrations as determined by the reaction mechanism.

Therefore, while slightly modified reaction rates as a result of turbulence-chemistry interaction is expected to have an influence on the magnitude and location of the eigenvalue associated with the explosive mode, qualitative identification of the explosive mode and its role in inferring the flame stabilization mechanism is expected to remain unaltered.

Additionally, since the grid used in the current study is not necessarily refined to resolve the flame thickness, turbulence-chemistry interaction is also affected by the subgrid scale turbulence effects. Once again, its influence on the CEM analysis is expected to be confined to the accuracy of prediction of the magnitude of the explosive mode and instantaneous spatial locations of the explosive region as a result of the turbulence-chemistry interaction, but it does not alter the mechanism of ignition or flame stabilization.

\section{Conclusion}
This work demonstrates the capability of the Improved Delayed Detached Eddy Simulation (IDDES) model, combined with a finite rate chemistry (FRC) approach, to accurately capture the complex, unsteady flow dynamics in a radical-farming scramjet engine. By employing a novel, fully coupled modeling strategy encompassing all engine components, from inlet to nozzle, the study establishes a robust framework for comprehensive full-scale engine analysis.\\

A key achievement of this work was the accurate prediction of auto-ignition location, a challenge that had previously been a source of error in steady-state simulations. The IDDES model, when used with low-dissipation numerical methods, enabled a high-fidelity representation of the flow, with computational schlieren images and OH concentration profiles showing strong agreement with experimental data. The simulation captured the turbulent mixing and combustion processes, with the majority of combustion occurring efficiently within the isolator and combustor regions. Wall pressure distributions were more stable and exhibited fewer fluctuations at an equivalence ratio of 0.5, while higher equivalence ratios of 0.8 and above showed greater variability and steeper gradients, reflecting greater heat release and instabilities captured by the simulations. Further, Takeno flame index analysis  identified multiple combustion regimes within the combustor, with ignition occurring in the partially premixed regime. Varying equivalence ratios revealed a clear link between increased fuel concentration and more intense combustion, resulting in higher temperatures and faster fuel consumption, driven primarily by enhanced compressibility and turbulence at elevated equivalence ratios. Chemical Explosive Mode Analysis (CEMA) identified chemically sensitive regions matching experimentally observed hot spots, providing insights into autoignition locations and flame stabilization mechanisms. The detailed 12-species, 27-reaction kinetics mechanism \cite{Burke2011} was essential to capture intermediate radicals such as \ce{HO2} and \ce{H2O2} critical to autoignition and flame stability. Integrating this chemical model with high-resolution turbulence modeling provides a predictive framework for simulating thermo-chemical interactions vital to hypersonic propulsion.\\

Overall, the findings validate the IDDES approach as a powerful tool for investigating scramjet propulsion systems and offer a framework for tackling the significant challenges of future air-breathing propulsion. Future efforts will focus on implementing Large Eddy Simulation (LES) and investigating subgrid-scale model sensitivities to further improve predictive accuracy. Future work should also focus on applying this validated framework to more complex engine geometries or different operational conditions to further enhance the design and optimization of hypersonic propulsion systems.
\label{sec:conclusions}

\section*{Appendix}
An Appendix, if needed, should appear before the acknowledgments.

\section*{Acknowledgments}
This research was supported by the DEVCOM Army Research Laboratory under grants W911NF-19-1-0225 and W911NF-22-2-0058. Additional support was provided in part by the 2025 Department of Defense High Performance Computing Internship Program (HIP) Workforce Development Mentorship Grant, through resources from the Department of Defense High Performance Computing Modernization Program (HPCMP). This work was administered by the Oak Ridge Institute for Science and Education (ORISE) under an interagency agreement between the U.S. Department of Energy (DOE) and the Department of Defense (DoD). ORISE is managed by ORAU under DOE contract number DE-SC0014664.

The authors gratefully acknowledge the HPCMP resources and support provided by the Department of Defense Supercomputing Resource Center (DSRC) as part of the 2022 Frontier Project, Large-Scale Integrated Simulations of Transient Aerothermodynamics in Gas Turbine Engines.

The views and conclusions contained in this document are those of the authors and should not be interpreted as representing the official policies or positions, either expressed or implied, of the DEVCOM Army Research Laboratory or the U.S. Government. The U.S. Government is authorized to reproduce and distribute reprints for governmental purposes notwithstanding any copyright notation herein.

\bibliography{sample}

\end{document}